\documentclass[12pt]{article}

\usepackage{amssymb} 
\usepackage{amsmath}

\newtheorem{theorem}{Theorem}[section]
\newtheorem{lemma}[theorem]{Lemma}
\newtheorem{proposition}[theorem]{Proposition}
\newtheorem{corollary}[theorem]{Corollary}

\newtheorem{remark}[theorem]{Remark}

\newcommand{\bull}{\mbox{$\;\;\;$\vrule height .9ex width .8ex depth -.1ex}}
\newcommand{\qed}{$\;\;\;\Box$}
\newenvironment{proof}{\par\smallbreak\noindent{\bf Proof.~}}
{\unskip\nobreak\hfill \bull \par\medbreak}

{\qed \par\smallbreak}
\newcounter{claim}[theorem]
\renewcommand{\theclaim}{\thetheorem.\arabic{claim}}
{\par\smallskip\par}

\newcommand{\CASE}[2]%
{\par\medskip\par\noindent{\it Case #1:\/ #2}\par\smallskip\par\noindent}
\newenvironment{venumerate}%
{\begin{enumerate}}%
{\end{enumerate}}
\newcommand{\refeq}[1]{(\ref{#1})}

\newcommand{\setdef}[2]{\left\{ \hspace{0.5mm} #1 :
\hspace{0.5mm} #2 \right\}}
\newcommand{\msetdef}[2]{\left\{\!\!\left\{ \hspace{0.5mm} #1 :
\hspace{0.5mm} #2 \right\}\!\!\right\}}

\newcommand{\of}[1]{\left( #1 \right)}

\newcommand{\diam}[1]{\mathit{diam}\,(#1)}

\newcommand{\baru}{{\bar u}}
\newcommand{\barv}{{\bar v}}
\newcommand{\isotype}[1]{{\textup{tp}(#1)}}
\newcommand{\wl}[3]{W^{#1,#2}(#3)}

\newcommand{\cd}[2]{{\scriptstyle c}D^{#1}(#2)}
\newcommand{\game}{\mbox{\sc Ehr}}
\newcommand{\EF}{Ehrenfeucht-Fra\"\i{}ss\'{e}}
\newcommand{\WL}[1]{$#1$-dim WL}
\newcommand{\tlt}[1]{\par\medskip{\sc #1}\par\smallskip\par\noindent}
\newcommand{\subtlt}[1]{\par\medskip{\it #1}}

\newcommand{\tc}[1]{\mbox{\rm TC$^{#1}$}}
\newcommand{\nc}[1]{\mbox{\rm NC$^{#1}$}}
\newcommand{\ac}[1]{\mbox{\rm AC$^{#1}$}}
\newcommand{\flap}{\odot}
\newcommand{\betw}{\ominus}

\newcommand{\mso}{\{\!\!\{}
\newcommand{\msc}{\}\!\!\}}

\title{Testing Graph Isomorphism in Parallel\\ 
by Playing a Game}

\author{Martin Grohe\ \ and\ \ Oleg Verbitsky%
\thanks{Supported by an Alexander von Humboldt fellowship.}\\[4mm]
Institut f\"ur Informatik\\
Humboldt Universit\"at zu Berlin, D-10099 Berlin}

\date{}

\begin{document}

\sloppy

\maketitle

\begin{abstract}
  Our starting point is the observation that if graphs in a class $C$ have low
  descriptive complexity, then the isomorphism problem for $C$ is solvable by
  a fast parallel algorithm.  More precisely, we prove that if every graph in
  $C$ is definable in a finite-variable first order logic with counting
  quantifiers within logarithmic quantifier depth, then Graph Isomorphism for
  $C$ is in $\tc1\subseteq\nc2$. If no counting quantifiers are needed, then
  Graph Isomorphism for $C$ is even in $\ac1$. The definability conditions can
  be checked by designing a winning strategy for suitable
  Ehrenfeucht-Fra\"\i{}ss\'{e} games with a logarithmic number of rounds.  The
  parallel isomorphism algorithm this approach yields is a simple
  combinatorial algorithm known as the Weisfeiler-Lehman (WL) algorithm.
  
  Using this approach, we prove that isomorphism of graphs of bounded
  treewidth is testable in \tc1, answering an open question from \cite{Cha}.
  Furthermore, we obtain an \ac1 algorithm for testing isomorphism of rotation
  systems (combinatorial specifications of graph embeddings). The \ac1 upper
  bound was known before, but the fact that this bound can be achieved by the
  simple WL algorithm is new. Combined with other known results, it also
  yields a new \ac1 isomorphism algorithm for planar graphs.
\end{abstract}

\section{Introduction}

\subsection{The Graph Isomorphism problem}
An isomorphism between two graphs $G$ and $H$ is a 1-to-1 correspondence
between their vertex sets $V(G)$ and $V(H)$ that relates edges to edges and
non-edges to non-edges. Two graphs are isomorphic if there exists an
isomorphism between them. \emph{Graph Isomorphism (GI)} is the problem of
recognizing if two given graphs are isomorphic. The problem plays a prominent
role in complexity theory as one of the few natural problems in NP that are
neither known to be NP-complete nor known to be in polynomial time. There are
good reasons to believe that GI is not NP-complete; most strikingly, this
would imply a collapse of the polynomial hierarchy \cite{BHZ,Sch}.  The best
known graph isomorphism algorithm due to Babai, Luks, and Zemplyachenko
\cite{bab81,babluk83} takes time $O(2^{\sqrt{n\log n}})$, where $n$ denotes
the number of vertices in the input graphs. The strongest known hardness
result \cite{Tor} says that GI is hard for DET, which is a subclass of \nc2.
The complexity status of GI is determined precisely only if the problem is
restricted to trees: For trees GI is LOGSPACE-complete \cite{Lin,JKMT}.

However, there are many natural classes of graphs such that the restriction of
GI to input graphs from these classes is in polynomial time. These include
planar graphs~\cite{HTa}, graphs of bounded
genus~\cite{FMa,Mil}, graphs of bounded treewidth~\cite{Bod}, graphs with
excluded minors~\cite{pon88}, graphs of bounded degree~\cite{Luk}, and graphs
of bounded eigenvalue multiplicity~\cite{babgrimou82}.
Linear-time algorithms are designed for planar graphs \cite{hopwon74}
and for graphs of treewidth at most three~\cite{APr}.
Here we are interested in classes of graphs for which the isomorphism problem
is solvable by a fast (i.e., polylogarithmic) parallel algorithm.
Recall the class NC and its refinements (e.g., \cite[Theorem 27.13]{ALR}):
$$
\nc{}=\textstyle\bigcup_i\nc i\mbox{\ \ and\ \ }
\nc i\subseteq\ac i\subseteq\tc i\subseteq\nc{i+1},
$$
where \nc i\ consists of functions computable by circuits of polynomial
size and depth $O(\log^in)$, \ac i\ is an analog for circuits with unbounded
fan-in, and \tc i\ is an extension of \ac i\ allowing threshold gates.  As
well known \cite{KRa}, \ac i\ consists of exactly those functions computable
by a CRCW PRAM with polynomially many processors in time $O(\log^in)$.  Miller
and Reif \cite{MRe} design an \ac1 algorithm for planar graph isomorphism
and isomorphism of rotation systems, which are combinatorial specifications of
graph embeddings (see \cite[Section 3.2]{MTh}). 
Chandrasekharan \cite{Cha} designs an \ac2\ isomorphism
algorithm for $k$-trees, a proper subclass of graphs of treewidth $k$, and
asks if there is an NC algorithm for the whole class of graphs with
treewidth $k$ (see also~\cite{DSS}).

We answer this question in affirmatively by showing that isomorphism of graphs
with bounded treewidth is in \tc1 (see Corollary~\ref{cor:btw}). Furthermore,
we obtain a new \ac1-algorithm for testing isomorphism of rotation systems
(see Corollary~\ref{cor:map}), which by techniques due to Miller and
Reif~\cite{MRe} also yields a new \ac1 isomorphism algorithm for planar graphs
(see Corollary~\ref{cor:planar}).

Remarkably, the algorithm we employ for both graphs of bounded treewidth
and rotation systems is a simple combinatorial algorithm that is actually
known since the late 1960s from the work of Weisfeiler and Lehman. This is
what we believe makes our result on rotation systems worthwhile, even though
in this case the \ac1 upper bound was known before.

\subsection{The multidimensional Weisfeiler-Lehman algorithm}\label{ss:wl}
For the history of this approach to GI we refer the reader to
\cite{Bab,CFI,evdkarpon99,evdpon99}.  We will abbreviate \emph{$k$-dimensional
  Weisfeiler-Lehman algorithm} by \emph{\WL k}. The \WL1\ is commonly known as
\emph{canonical labeling} or \emph{color refinement algorithm}. It proceeds in
rounds; in each round a coloring of the vertices of the input graphs $G$ and
$H$ is defined, which refines the coloring of the previous round. The initial
coloring $C^0$ is uniform, say, $C^0(v)=1$ for all vertices $v\in V(G)\cup
V(H)$. In the $(i+1)$st round, the color $C^{i+1}(v)$ is defined to be a pair
consisting of the preceding color $C^{i-1}(v)$ and the multiset of colors
$C^{i-1}(u)$ for all $u$ adjacent to $v$. For example, $C^1(v)=C^1(w)$ iff $v$
and $w$ have the same degree.  To keep the color encoding short, after each
round the colors are renamed (we never need more than $2n$ color names). As the
coloring is refined in each round, it stabilizes after at most $2n$ rounds,
that is, no further refinement occurs. The algorithm stops as soon as this
happens. If the multiset of colors of the vertices of $G$ is distinct from the multiset
of colors of the vertices of $H$, the algorithms reports that the graphs are
not isomorphic; otherwise, it declares them to be isomorphic. Clearly, this
algorithm is not correct. It may report false positives, for example, if both
input graphs are regular with the same vertex degree, the coloring stabilizes
after the first round, and all vertices of both graphs have the same color.
It is known that the \WL1\ works correctly for almost all $G$ (and every $H$) \cite{BES} 
and for all trees~\cite{AHU}. 

Following the same idea, the $k$-dimensional version iteratively refines a
coloring of $V(G)^k\cup V(H)^k$. The initial coloring of a $k$-tuple $\barv$
is the isomorphism type of the subgraph induced by the vertices in $\barv$
(viewed as a labeled graph where each vertex is labeled by the positions in the
tuple where it occurs). The refinement step takes into account the colors of
all neighbors of $\barv$ in the Hamming metric (see details in Section
\ref{s:wl}).  Color stabilization is now reached in
$
r<2n^k
$
rounds. The \WL k\ is polynomial-time for each constant $k$.
In 1990, Cai, F\"urer, and Immerman \cite{CFI} proved a striking negative
result: For any sublinear dimension $k=o(n)$, the \WL k\ does not work
correctly even on graphs of vertex degree 3. Nevertheless, later it
was realized that a constant-dimensional WL is still applicable to particular
classes of graphs, including planar graphs \cite{Gro1}, graphs
of bounded genus \cite{Gro2}, and graphs of bounded treewidth~\cite{GMa}.

We show that the \WL k\ admits a natural parallelization such that the number
of parallel processors and the running time are closely related to $n^k$ and
$r$, respectively, where $r$ denotes the number of rounds performed by the
algorithm.  Previous work never used any better bound on $r$ than the trivial
$r<2n^k$, which was good enough to keep the running time polynomially
bounded.  In view of a possibility that $r$ can be much smaller, 
we show that the $r$-round \WL k\ can be implemented
on a logspace uniform family of circuits with gates of unbounded fan-in and
threshold gates (such circuits are used to define the TC hierarchy) of depth
$O(r)$ and size $O(r\cdot n^{3k})$. It follows that if for a class of
graphs $C$ there is a constant $k$ such that for all $G,H\in C$ the \WL k\ 
in $O(\log n)$ rounds 
correctly decides if $G$ and $H$ are isomorphic or not, then there is a \tc1
algorithm deciding GI on $C$. We also prove a version of these results for a
related algorithm we call the \emph{count-free WL algorithm} that places GI on
suitable classes $C$ into \ac1.

\subsection{Descriptive complexity of graphs}
To prove that the \WL k\ correctly decides isomorphism of graphs
from a certain class $C$ in a logarithmic number of rounds, we
exploit a close relationship between the WL algorithm and the descriptive
complexity of graphs, which was discovered in \cite{CFI}: The $r$-round
\WL k\ correctly decides if two graphs $G$ and $H$ are isomorphic
in at most $r$ rounds if and only if $G$ and $H$ are distinguishable
in the $(k+1)$-variable first order logic with counting quantifiers in the
language of graphs by a sentence of quantifier depth $r$.
(In)distinguishability of two graphs in various logics can be characterized in
terms of so-called Ehrenfeucht-Fra\"\i{}ss\'{e} games. The appropriate game
here is the counting version of the $r$-round $k$-pebble game (see Section
\ref{s:ehr}). The equivalences between correctness of the $r$-round \WL k, the
logical indistinguishability result, and its game characterization reduces the
design of a \tc1 isomorphism algorithm on $C$ to design of winning strategies
in the $O(\log n)$-round $k$-pebble counting game on graphs from the class
$C$, where $k$ is a constant. 
Similarly, the design of an \ac1 isomorphism algorithm on $C$ can be
reduced to the design of winning strategies in the $O(\log n)$-round
$k$-pebble game (without counting) on graphs from the class $C$.

Our results on the descriptive complexity of graphs are actually slightly stronger 
than it is needed for algorithmic applications:
They give $O(\log
n)$ upper bounds on the quantifier depth of sentences in the $k$-variable
first-order logic (with or without counting) required to distinguish a graph
$G$ from all other graphs. For graphs of
treewidth at most $k$, we obtain an $O(k\cdot\log n)$ upper bound on the
quantifier depth of sentences in the $(4k+4)$-variable first-order logic with
counting (see Theorem~\ref{thm:btw}). For rotation systems, we obtain an
$O(\log n)$ upper bound  on the
quantifier depth of sentences in the $5$-variable first-order logic without
counting (see Theorem~\ref{thm:map}). The proofs are based on 
an analysis of Ehrenfeucht-Fra\"\i{}ss\'{e} games.

Various aspects of descriptive complexity of graphs have recently been
investigated in \cite{BFL+,KPSV,PSV,PVV,Ver} with focus on the minimum
quantifier depth of a first order sentence defining a graph. In particular, a comprehensive analysis
of the definability of trees in first order logic is carried out in
\cite{BFL+,PSV,Ver}.
Here we extend it to the definability 
of graphs with bounded treewidth in first order logic with counting. 
Notice a fact that makes our results on
descriptive complexity potentially stronger (and harder to prove):
We are constrained by the condition that a defining sentence must be
in a finite-variable logic.

\subsection{Organization of the paper}
In Section \ref{s:log} we give relevant definitions from descriptive complexity
of graphs. The Weisfeiler-Lehman algorithm is treated in Section \ref{s:wl}.
Section \ref{s:graphth} contains some graph-theoretic preliminaries.
Section \ref{s:ehr} is devoted to the  Ehrenfeucht-Fra\"\i{}ss\'{e} game.
We prove our results about graphs of bounded treewidth in Section \ref{s:btw}
and about rotation systems and planar graphs in Section~\ref{s:maps}.

\section{Logical depth of a graph}\label{s:log}

Let $\Phi$ be a first order sentence about a graph in the language
of the adjacency and the equality relations. We say that $\Phi$
\emph{distinguishes} a graph $G$ from a graph $H$ if $\Phi$ is true
on $G$ but false on $H$. We say that $\Phi$
\emph{defines} $G$ if $\Phi$ is true on $G$ and false on any graph
non-isomorphic to $G$. The quantifier rank of $\Phi$ is the maximum number of nested
quantifiers in $\Phi$. The \emph{logical depth} of a graph $G$, denoted by $D(G)$, 
is the minimum quantifier depth of $\Phi$ defining $G$.

The \emph{$k$-variable logic} is the fragment of first order logic where
usage of only $k$ variables is allowed. 
If we restrict defining sentences to the $k$-variable logic, this variant
of the logical depth of $G$ is denoted by $D^k(G)$.
We have 
\begin{equation}\label{eq:ddd}
D^k(G)=\max\setdef{D^k(G,H)}{H\not\cong G},
\end{equation}
where $D^k(G,H)$ denotes the minimum quantifier depth of a $k$-variable sentence
distinguishing $G$ from $H$.
This equality easily follows from the fact that, for each $r$, 
there are only finitely many pairwise inequivalent first order sentences
about graphs of quantifier depth at most $r$. It is assumed that
$D^k(G)=\infty$ (resp.\ $D^k(G,H)=\infty$) if the $k$-variable logic is too
weak to define $G$ (resp.\ to distinguish $G$ from $H$).

Furthermore, let $\cd kG$ (resp.\ $\cd k{G,H}$) denote the variant of $D^k(G)$
(resp.\ $D^k(G,H)$) for the first order logic
with \emph{counting quantifiers} where we allow expressions of the type
$\exists^m\Psi$ to say that there are at least $m$ vertices with property $\Psi$
(such a quantifier contributes 1 in the quantifier depth irrespective of $m$).
Similarly to \refeq{eq:ddd} we have
\begin{equation}\label{eq:dddc}
\cd kG=\max\setdef{\cd k{G,H}}{H\not\cong G}.
\end{equation}

\section{The \WL k\ as a parallel algorithm}\label{s:wl}

Let $k\ge2$.
Given an ordered $k$-tuple of vertices $\baru=(u_1,\ldots,u_k)\in V(G)^k$, we
define the \emph{isomorphism type} of $\bar u$ to be the pair
\[
\isotype\baru=\Big(\big\{(i,j)\in[k]^2\mid u_i=u_j\}\big\},\{(i,j)\in[k]^2\mid \{u_i,u_j\}\in
E(G)\big\}\Big),
\]
where $[k]$ denotes the set $\{1,\ldots,k\}$.
If $w\in V(G)$ and $i\le k$, we let $\baru^{i,w}$ denote the result
of substituting $w$ in place of $u_i$ in $\baru$. 

The \emph{$r$-round $k$-dimensional Weisfeiler-Lehman algorithm ($r$-round \WL k)}
takes as an input two graphs $G$ and $H$ and purports to decide
if $G\cong H$. The algorithm performs 
the following operations with the set $V(G)^k\cup V(H)^k$.

\tlt{Initial coloring.}
The algorithm assigns each $\baru\in V(G)^k\cup V(H)^k$ color
$\wl k0\baru=\isotype\baru$ (in a suitable encoding).

\tlt{Color refinement step.}
In the $i$-th round each $\baru\in V(G)^k$ is assigned color
$$
\wl ki\baru=\of{\wl k{i-1}\baru,
\msetdef{(\wl k{i-1}{\baru^{1,w}},\ldots,\wl k{i-1}{\baru^{k,w}})}
{w\in V(G)}}
$$
and similarly with each $\baru\in V(H)^k$.

Here $\mso\ldots\msc$ denotes a multiset. In a variant of the algorithm,
which will be referred to as the \emph{count-free version}, this is a set.

\tlt{Computing an output.}
The algorithm reports that
$G\not\cong H$ if
\begin{equation}\label{eq:decision}
\msetdef{\wl kr{\baru}}{\baru\in V(G)^k}\ne
\msetdef{\wl kr{\baru}}{\baru\in V(H)^k}.
\end{equation}
and that $G\cong H$ otherwise.

In the above description we skipped an important implementation detail.
Denote the minimum length of the code of $\wl ki\baru$ over all $\baru$
by $L(i)$. As easily seen, for any natural encoding we should expect that
$L(i)\ge(k+1)L(i-1)$. To prevent increasing $L(i)$ at the exponential rate,
before every refinement step we arrange colors of all $k$-tuples 
of $V(G)^k\cup V(H)^k$ in the lexicographic order and replace each color with 
its number.

As easily seen, if $\phi$ is an isomorphism from $G$ to $H$, then for all $k$, $i$,
and $\baru\in V(G)^k$ we have $\wl ki\baru=\wl ki{\phi(\baru)}$.
This shows that for the isomorphic input graphs the output is always correct.
We say that the $r$-round \WL k\ \emph{works correctly for a graph $G$}
if its output is correct on all input pairs $(G,H)$.

\begin{proposition}{\bf(Cai-F\"urer-Immerman \cite{CFI})}\label{prop:WLEF}
\begin{venumerate}
\item
The $r$-round \WL k\ works correctly for a graph $G$ iff $r\ge\cd{k+1}G$.
\item
The count-free $r$-round \WL k\ works correctly for a graph $G$ iff $r\ge D^{k+1}(G)$.
\end{venumerate}
\end{proposition}

Let us call a circuit with gates of unbounded fan in an \emph{AC-circuit}. If,
in addition, the circuit contains threshold gates, then we call it a \emph{TC-circuit}.

\begin{theorem}\label{thm:TC}
  Let $k\ge2$ be a constant and $r=r(n)$ a function, where $n$ denotes the
  order of the input graphs.
  \begin{venumerate}
  \item The $r$-round \WL k\ can be implemented by a logspace uniform family
    of TC-circuits of depth $O(r)$ and size $O(r\cdot n^{3k})$.
  \item The $r$-round count-free \WL k\ can be implemented by a logspace
    uniform family of AC-circuits of depth $O(r)$ and size $O(r\cdot
    n^{3k})$.
  \end{venumerate}
\end{theorem}

\begin{proof}
 Let $N=2n^k$. We fix a 1-to-1 correspondence between
  numbers in $[n^k]$ and tuples in $V(G)^k$ and numbers in
  $\{n^k+1,\ldots,N\}$ and tuples in $V(H)^k$. For every $a\in [N]$, let
  $\baru(a)$ denote the tuple in $V(G)^k\cup V(H)^k$ corresponding to $a$.

  We shall construct a circuit that consists of $r+2$ layers, where each layer
  is a constant depth circuit. Layer $0$ computes the initial coloring. For
  $1\le \ell\le r$, layer $\ell$ is used to refine the coloring obtained in the
  previous layer as described in the refinement step of the WL algorithm.
  Finally, layer $(r+1)$ is used to compute the output.   
  Depending on which version of the algorithm we use, the circuit will be a
  TC-circuit or an AC-circuit.

  For $0\le \ell\le r$,
  layer $\ell$ will have outputs $X_\ell(a,c)$ for all $a,c\in[N]$ such
  that (for all inputs) for every $a\in[N]$ there is exactly one $c\in[N]$
  such that $X_\ell(a,c)$ evaluates to $1$; we think of this $c$ as the name
  of the color of $\baru(a)$ after the $\ell$th refinement step.  
  
  To define layer $0$, note that the isomorphism type $\isotype{\bar u}$ of a
  tuple $\bar u$ can be described by a bitstring of length $2k^2$. On the
  bottom of layer $0$, there is a bounded depth AC-circuit with $2k^2\cdot N$
  output gates that computes this bitstring for every $a\in[N]$. For every
  $a\in[N]$ and $j\in[2k^2]$, let $Z_0(a,j)$ be the output of this
  circuit that computes the $j$th bit of the bitstring encoding $\isotype{\bar
    u(a)}$.  For all
  $a,b\le[N]$, we let
  \[
  Y_0(a,b)=\bigwedge_{i=1}^{2k^2}\big(Z_0(a,i)\leftrightarrow Z_0(b,i)\big),
  \]
  where $\big(Z_0(a,i)\leftrightarrow Z_0(b,i)\big)$ abbreviates
  $\big((Z_0(a,i)\wedge Z_0(b,i))\vee(\neg Z_0(a,i)\wedge \neg
  Z_0(b,i))\big)$.
  Then $Y_0(a,b)=1\iff\isotype{\bar u(a)}=\isotype{\bar u(b)}$. Now we define 
  $X_0(a,c)$ by the formula
  \begin{equation}
    \label{eq:ref0}
  Y_0(a,c)\wedge\bigwedge_{d=1}^{c-1}Y_0(a,d).
  \end{equation}
  Thus the name of the color of $\baru(a)$ is simply the index $c$ of the
  first tuple that has the same isomorphism type as $\baru(a)$. We can view
  the formulas defining the $Y_0(a,b)$ and the $X_0(a,c)$ as constant depth
  circuits on top of the constant depth circuit defining the $Z_0(a,j)$ This
  completes the definition of layer $0$.
  
  Now let $\ell\in[r]$, and assume that layers $0,\ldots,\ell-1$ have already
  been defined. We first want to define a circuit with outputs
  $Y_\ell(a,b)$ such that $Y_\ell(a,b)=1$ if and only if $\baru(a)$ and
  $\baru(b)$ have the same color after round $\ell$. For every $a\in[N]$, let
  $c_{\ell-1}(a)$ be the name of the color of $\bar u(a)$ after round
  $\ell-1$, that is, $c_{\ell-1}(a)$ is the unique $c$ such that
  $X_{\ell-1}(a,c)$ evaluates to $1$. For $j\in[k]$ and $v\in V(G)$, let
  $a^{jv}$ be the index of the tuple obtained from $\baru(a)$ by replacing the
  $j$-the position by $v$. Thus we have $\baru(a^{jv})=\baru(a)^{jv}$.
  
  Now we have to
  distinguish between the counting and the count-free algorithm. 
  We consider the counting algorithm first. Recall that, for $a,b\in[N]$,
  $\baru(a)$ and $\baru(b)$ have the same color after round $\ell$ if and only
  if 
  \begin{equation}
    \label{eq:ref1}
      c_{\ell-1}(a)=c_{\ell-1}(b)
  \end{equation}
  and 
  \begin{equation}
    \label{eq:ref2}
    \begin{array}{cl}
        &\mso (c_{\ell-1}(a^{1v}),\ldots,c_{\ell-1}(a^{kv}))\mid v\in V(G)\msc\\
  =&
  \mso (c_{\ell-1}(b^{1w}),\ldots,c_{\ell-1}(b^{kw}))\mid w\in V(H)\msc.
  \end{array}
  \end{equation}
  Condition \eqref{eq:ref1} can be defined by the formula
  \begin{equation}
    \label{eq:ref4}
  \bigwedge_{c\in[N]}\big(X_{\ell-1}(a,c)\leftrightarrow X_{\ell-1}(b,c)\big).
  \end{equation}
  To express condition \eqref{eq:ref2}, for all $v,v'\in V(G)$, let
  $\phi(v,v')$ be the following formula expressing that 
  $
  (c_{\ell-1}(a^{1v}),\ldots,c_{\ell-1}(a^{kv}))=(c_{\ell-1}(a^{1v'}),\ldots,c_{\ell-1}(a^{kv'})):
  $
  \[
  \phi(v,w)=\bigwedge_{j\in[k]}\bigwedge_{c\in[N]}
  \big(X_{\ell-1}(a^{jv'},c)\leftrightarrow X_{\ell-1}(a^{jv},c)\big)
  \]
  Similarly, for all $v\in V(G),w\in V(H)$ we can define a formula $\psi(v,w)$
  expressing that 
    $
  (c_{\ell-1}(a^{1v}),\ldots,c_{\ell-1}(a^{kv}))=(c_{\ell-1}(b^{1w}),\ldots,c_{\ell-1}(b^{kw})).
  $
  Then condition \eqref{eq:ref2} can
  be defined by the formula
  \begin{equation}
    \label{eq:ref5}
    \bigwedge_{v\in V(G)}
    \left(
      \left(
        \sum_{v'\in V(G)}\phi(v,v')
      \right)
      =
      \left(
        \sum_{w\in V(H)}\psi(v,w)
      \right)
    \right).
  \end{equation}
  The conjunction of the formulas \eqref{eq:ref4} and \eqref{eq:ref5} yields a
  definition of $Y_\ell(a,b)$ that can easily be turned into a constant depth
  TC-circuit.

  Now let us consider the count-free version of the algorithm. In this case,
  condition \eqref{eq:ref2} has to be replaced by 
    \begin{equation}
    \label{eq:ref6}
    \begin{array}{cl}
        &\{ (c_{\ell-1}(a^{1v}),\ldots,c_{\ell-1}(a^{kv}))\mid v\in V(G)\}\\
  =&
  \{ (c_{\ell-1}(b^{1w}),\ldots,c_{\ell-1}(b^{kw}))\mid w\in V(H)\}.
  \end{array}
  \end{equation}
  which can be expressed by the formula
  \begin{equation}
    \label{eq:ref7}
    \bigwedge_{v\in V(G)}\bigvee_{w\in V(H)}\psi(v,w)\wedge\bigwedge_{w\in
    V(H)}\bigvee_{v\in V(G)}\psi(v,w).
  \end{equation}
  The conjunction of the formulas \eqref{eq:ref4} and \eqref{eq:ref7} yields a
  definition of $Y_\ell(a,b)$ for the count-free version that can easily be
  turned into a constant depth AC-circuit.
  
  To complete the definition of layer $\ell$, we proceed as in \eqref{eq:ref0}
  for layer $0$. That is, we define the outputs $X_\ell(a,c)$ of layer $\ell$
  by 
  \[
  Y_\ell(a,c)\wedge\bigwedge_{d=1}^{c-1}Y_\ell(a,d).
  \]
  This completes the definition of layer $\ell$.

  Finally, the output of the overall circuit is defined in layer $(r+1)$ by
  the formula
  \[
  \bigwedge_{c\in[N]}\left(\sum_{a=1}^{n^k}X_r(a,c)=\sum_{b=n^k+1}^NX_r(b,c)
  \right)
  \]
  for the counting version and the formula
  \[
  \bigwedge_{c\in[N]}\left(\bigvee_{a=1}^{n^k}X_r(a,c)\leftrightarrow\bigvee_{b=n^k+1}^NX_r(b,c)
  \right)
  \]
  in the count-free version.
\end{proof}

The following corollary states the most important application of the previous
theorem for us:

\begin{corollary}\label{cor:TC}
  Let $k\ge2$ be a constant.
  \begin{venumerate}
  \item
    Let $C$ be a class of graphs $G$ with
    $\cd {k+1}G=O(\log n)$. Then Graph Isomorphism for $C$ is in \tc1.
  \item
    Let $C$ be a class of graphs $G$ with $D^{k+1}(G)=O(\log n)$.
    Then Graph Isomorphism for $C$ is in \ac1.
  \end{venumerate}
\end{corollary}

\begin{remark}\rm
The Weisfeiler-Lehman algorithm naturally generalizes from graphs
to an arbitrary class of structures over a fixed vocabulary.
It costs no extra efforts to extend Theorem \ref{thm:TC}
as well as Corollary \ref{cor:TC} in the general situation
and we will use this in Section~\ref{s:maps}.
\end{remark}

\section{Graph-theoretic preliminaries}\label{s:graphth}

The distance between vertices $u$ and $v$ in a graph $G$ is denoted by $d(u,v)$.
If $u$ and $v$ are in different connected components, we set $d(u,v)=\infty$.
The diameter of $G$ is defined by $\diam G=\max\setdef{d(u,v)}{u,v\in V(G)}$.
The set $\Gamma(v)=\setdef{u}{d(u,v)=1}$ is called the \emph{neighborhood}
of a vertex $v$ in $G$.
Let $X\subset V(G)$. The subgraph induced by $G$ on $X$ is denoted by $G[X]$.
We denote $G\setminus X=G[V(G)\setminus X]$, which is the result of removal of all
vertices in $X$ from $G$. 
We call the vertex set of a connected component of
$G\setminus X$ \emph{a flap of $G\setminus X$}. We call $X$ 
\emph{a separator of $G$} if every flap of $G\setminus X$ has at most
$|V(G)|/2$ vertices.

A \emph{tree decomposition} of a graph $G$ is a tree $T$ and a family
$\{X_i\}_{i\in V(T)}$ of sets $X_i\subseteq V(G)$, called \emph{bags},
such that the union of all bags covers all $V(G)$, every edge of $G$
is contained in at least one bag, and we have $X_i\cap X_j\subseteq X_l$
whenever $l$ lies on the path from $i$ to $j$ in $T$.

We will use the following three properties of tree decompositions.
The first two can be found in \cite[Lemmas 12.3.1--2]{Die}, 
the third is due to~\cite{RSe}.

\begin{proposition}\label{prop:treedec}
Let $(T,\{X_i\}_{i\in V(T)})$ be a tree decomposition of $G$.
\begin{venumerate}
\item
Let $Z\subseteq V(G)$. Then  $(T,\{X_i\cap Z\}_{i\in V(T)})$
is a tree decomposition of $G[Z]$.
\item
Suppose that $l$ lies on the path from $i$ to $j$ in $T$.
Then every path from $X_i$ to $X_j$ in $G$ visits $X_l$.
\item
There is a bag $X_i$ that is a separator of~$G$.
\end{venumerate}
\end{proposition}
The \emph{width} of the decomposition is $\max|X_i|-1$.
The \emph{treewidth} of $G$ is the minimum width of a tree decomposition of~$G$.

Now we introduce a non-standard notation specific to our purposes.
It will be convenient to regard it as a notation for
two binary operations over set of vertices.
Let $A\subset V(G)$ and $v\in V(G)\setminus A$. Then 
$A\flap v$ denotes the union of $A$ and the flap of $G\setminus A$ containing $v$.
Furthermore, let $A,C\subset V(G)$ be nonempty and disjoint.
Then $A\betw C$ is the union of $A$, $C$, and the set of all those vertices
$x\in V(G)\setminus(A\cup C)$ such that there are a path from $x$ to $A$ in 
$G\setminus C$ and a path from $x$ to $C$ in $G\setminus A$.

\section{ Ehrenfeucht-Fra\"\i{}ss\'{e} game}\label{s:ehr}

Let $G$ and $H$ be graphs with disjoint vertex sets.
The \emph{$r$-round $k$-pebble \EF\/ game on $G$ and $H$},
denoted by $\game_r^k(G,H)$, is played by
two players, Spoiler and Duplicator, with $k$ pairwise distinct
pebbles $p_1,\ldots,p_k$, each given in duplicate. Spoiler starts the game.
A {\em round\/} consists of a move of Spoiler followed by a move of
Duplicator. At each move Spoiler takes a pebble, say $p_i$, selects one of
the graphs $G$ or $H$, and places $p_i$ on a vertex of this graph.
In response Duplicator should place the other copy of $p_i$ on a vertex
of the other graph. It is allowed to remove previously placed pebbles
to another vertex and place more than one pebble on the same vertex.

After each round of the game, for $1\le i\le k$ let $x_i$ (resp.\ $y_i$)
denote the vertex of $G$ (resp.\ $H$) occupied by $p_i$, irrespectively
of who of the players placed the pebble on this vertex. If $p_i$ is
off the board at this moment, $x_i$ and $y_i$ are undefined.
If after every of $r$ rounds the component-wise correspondence $(x_1,\ldots,x_k)$ to
$(y_1,\ldots,y_k)$ is a partial isomorphism from $G$ to $H$, this is
a win for Duplicator;  Otherwise the winner is Spoiler.

In the \emph{counting version} of the game,
the rules of $\game_r^k(G,G')$ are modified as follows.
A round now consists of two acts. First, Spoiler specifies a set of vertices
$A$ in one of the graphs. Duplicator responds with a set of vertices $B$
in the other graph so that $|B|=|A|$. Second, Spoiler places a pebble $p_i$
on a vertex $b\in B$. In response Duplicator has to place the other copy
of $p_i$ on a vertex $a\in A$.
We will say that Spoiler makes a \emph{composite move}.

\begin{proposition}\label{prop:game}{\bf (Immerman, Poizat, see \cite[Theorem 6.10]{Imm})}
\begin{venumerate}
\item
$D^k(G,H)$ equals the minimum $r$ such that Spoiler has a winning
strategy in $\game_r^k(G,H)$.
\item
$\cd k{G,H}$ equals the minimum $r$ such that Spoiler has a winning
strategy in the counting version of $\game_r^k(G,H)$.
\end{venumerate}
\end{proposition}

All the above definitions and statements have a perfect sense
for any kind of structures, in particular, for colored graphs.
On the vertex set of a \emph{colored graph} $G$ we have a certain
number of unary relations $C_i$, $i=1,2,\ldots$. If a vertex $v$ satisfies $C_i$,
we say that $v$ has color $i$. 
Of course, isomorphism and partial isomorphism of colored graphs
must respect the color relations. 
In Section \ref{s:maps} we deal with even more complicated structures,
with one binary and one ternary relations, and the notion of partial
isomorphism should be understood appropriately.

Throughout the paper $\log n$ denotes the binary logarithm.
Unless stated otherwise, $n$ will denote the number of vertices in a graph $G$.
In the rest of this section we develop elements of a \emph{fast strategy of Spoiler}
in the  Ehrenfeucht-Fra\"\i{}ss\'{e} game on graphs $G$ and $G'$. Referring to such a strategy
or saying that \emph{Spoiler wins fast} we will always mean that Spoiler is
able to win in the next $\log n+O(1)$ moves using only 3 pebbles, whatever
Duplicator's strategy. 
The following lemma provides us with a basic
primitive on which our strategy will be built.

\begin{lemma}\label{lem:halving}
Consider the game on graphs $G$ and $G'$. 
Let $u,v\in V(G)$, $u',v'\in V(G')$ and suppose that
$u,u'$ and as well $v,v'$ are under the same pebbles. 
Suppose also that $d(u,v)\ne d(u',v')$
and $d(u,v)\ne\infty$
(in particular, it is possible that $d(u',v')=\infty$).
Then Spoiler is able to win with 3 pebbles in $\lceil\log d(u,v)\rceil$ moves.
\end{lemma}

\begin{proof}
Spoiler uses the \emph{halving strategy}
(see \cite{Spe} for a detailed account).
\end{proof}

Consider the following configuration in the  Ehrenfeucht-Fra\"\i{}ss\'{e} game on graphs
$G$ and $G'$: A set of vertices $A$ and two vertices $v\notin A$ and $u$
are pebbled in $G$, while a set $A'$ and vertices $v'$ and $u'$ are pebbled in $G'$ 
correspondingly. Let $u\in A\flap v$ but $u'\notin A'\flap v'$.
Applying Lemma \ref{lem:halving} to graphs $G\setminus A$
and $G'\setminus A'$, we see that
Spoiler wins fast (operating with 3 pebbles but keeping all the pebbles on $A$
and $A'$).

Let now $u\notin A\flap v$ but $u'\in A'\flap v'$.
The symmetric argument only shows that Spoiler wins in less
than $\log\diam{G'}+1$ moves, whereas $\diam{G'}$ may be much larger than $n$.
However, Lemma \ref{lem:halving} obviously applies in the case that 
$\diam G\ne\diam{G'}$ and Spoiler wins fast anyway.

Assume that $\diam G=\diam{G'}$.
It follows that, if such $A,v,A',v'$ are pebbled and Spoiler decides
to move only inside $(A\flap v)\cup(A'\flap v')$, then Duplicator
cannot move outside for else Spoiler wins fast. In this situation
we say that Spoiler \emph{forces play in} $(A\flap v)\cup(A'\flap v')$
or \emph{restricts the game to} $G[A\flap v]$ and $G'[A'\flap v']$.

Similarly, if at some moment of the game we 
have two disjoint sets $A$ and $C$ of vertices
pebbled in $G$, then Spoiler can
force further play in $(A\betw C)\cup(A'\betw C')$, where 
$A',C'$ are the corresponding sets in~$G'$.

\section{Graphs of bounded treewidth}\label{s:btw}

\begin{theorem}\label{thm:btw}
If a graph $G$ on $n$ vertices has treewidth $k$,
then $$\cd{4k+4}G<2(k+1)\log n+8k+9.$$
\end{theorem}

On the account of Corollary \ref{cor:TC}.1 this has a consequence
for the computational complexity of Graph Isomorphism.

\begin{corollary}\label{cor:btw}
Let $k$ be a constant. The isomorphism problem for
the class of graphs with treewidth at most $k$ is in \tc1.
\end{corollary}

The rest of this section is devoted to the proof of Theorem \ref{thm:btw}.
It is based on Equality \refeq{eq:dddc} and Proposition \ref{prop:game}.
Let $G'\not\cong G$. We have to design a strategy for Spoiler
in the  Ehrenfeucht-Fra\"\i{}ss\'{e} game on $G$ and $G'$ allowing him to win with only
$4k+4$ pebbles in less than $2(k+1)\log n+8k+9$ 
moves, whatever Duplicator's strategy.
Fix $(T,\{X_s\}_{s\in V(T)})$, a depth-$k$ tree decomposition of $G$.

It is not hard to see that Spoiler can force play on $K$ and $K'$, 
some non-isomorphic components of $G$ and $G'$.
We hence can assume from the very beginning that $G$ and $G'$ are
connected. Moreover, we will assume that $\diam G=\diam{G'}$
because otherwise Spoiler has a fast win 
(in the sense discussed 
in Section \ref{s:ehr}).

We start with a high level description of the strategy.
The strategy splits the game into phases. Each phase can be
of two types, Type AB or Type ABC.
Whenever $X\subset V(G)$ consists of vertices pebbled in some moment of the game,
by default $X'$ will denote the set of vertices pebbled correspondingly in $G'$
and vice versa.
Saying that $G$ is \emph{colored according to the pebbling}, we mean
that every vertex which is currently pebbled by $p_j$ receives color $j$.

\tlt{Phase $i$ of type AB.}
Spoiler aims to ensure pebbling sets of vertices $A\subset V(G)$, $A'\subset V(G')$,
vertices $v\in V(G)\setminus A$, $v'\in V(G')\setminus A'$, and perhaps
sets of vertices $B\subset V(G)$, $B'\subset V(G')$ so that the following
conditions are met.

\begin{description}
\item[AB1]
Let $G_i=G[A_i\flap v_i]$ be colored according to the pebbling 
and $G'_i$ be defined similarly.
Then $G_i\not\cong G'_i$.
\item[AB2]
$|V(G_i)|\le|V(G_{i-1})|/2+k+1$ (we set $G_0=G$).
\item[AB3]
Both $G_i$ and $G'_i$ are connected. 
\item[AB4]
A set $B_i$ is pebbled if $|V(G_i)|>2k+2$, otherwise play comes to an endgame.
$B_i$ is a separator of $G_i$ and $B'_i\subset V(G'_i)$.
\item[AB5]
There are distinct $r,t\in V(T)$ such that $A_i\subseteq X_r$ and $B_i\subseteq X_t$.
\end{description}

\tlt{Phase $i$ of type ABC.}
Spoiler aims to ensure pebbling sets of vertices $A,C\subset V(G)$, $A',C'\subset V(G')$
so that $A\cap C=\emptyset$,
and perhaps sets $B\subset V(G)$, $B'\subset V(G')$ so that the following
conditions are met.

\begin{description}
\item[ABC1]
Let $G_i=G[A_i\betw C_i]$ be colored according to the pebbling
and $G'_i$ be defined similarly.
Then $G_i\not\cong G'_i$.
\item[ABC2]
$|V(G_i)|\le|V(G_{i-1})|/2+k+1$.
\item[ABC3]
Both $G_i$ and $G'_i$ are connected. 
\item[ABC4]
A set $B_i$ is pebbled if $|V(G_i)|>2k+2$, otherwise play comes to an endgame.
$B_i$ is a separator of $G_i$ and $B'_i\subset V(G'_i)$.
\item[ABC5]
There are pairwise distinct $r,s,t\in V(T)$ such that 
$s\in(\{r\}\flap t)\cap(\{t\}\flap r)$ and 
$A_i\subseteq X_r$, $B_i\subseteq X_s$, $C_i\subseteq X_t$.
\end{description}

In the next Phase $i+1$ Spoiler restricts the game to $G_i$ and $G'_i$,
keeping pebbles on $A_i\cup\{v_i\}$ (or $A_i\cup C_i$) until the new
$A_{i+1},v_{i+1}$ (or $A_{i+1},C_{i+1}$) are pebbled.
As soon as this is done, the pebbles on $(A_i\cup\{v_i\})\setminus(A_{i+1}\cup\{v_{i+1}\})$
(or $(A_i\cup C_i)\setminus(A_{i+1}\cup C_{i+1})$) can be released and reused by Spoiler
in further play.

\bigskip\bigskip\bigskip

\tlt{Endgame.}
Suppose it begins after Phase $l$.
We have $G_l\not\cong G'_l$ and the former graph has at most $2k+2$
vertices. Spoiler restricts the game to $G_l$ and $G'_l$.
If Duplicator agrees, Spoiler obviously wins in no more than $2k+2$ moves.
Once Duplicator moves outside, Spoiler has a fast win as explained in
Section \ref{s:ehr}, where \emph{fast} means less than 
$\log\diam{G_l}+2\le\log(k+1)+3$ moves.

\medskip

We now describe Spoiler's strategy in detail.
We adhere to an important convention:
Whenever referring to an induced subgraph of $G$ or $G'$, we suppose that
it is colored according to the existing pebbling.

\tlt{Phase 1.}
It has type AB.

\subtlt{Choice of $A_1$.}
Spoiler pebbles $A_1=X_p$, a bag which is a separator of $G$, see Proposition \ref{prop:treedec}.3.
We assume that Duplicator has not lost so far, i.e.,
the pebbling determines an isomorphism between $G[A_1]$ and $G'[A'_1]$
(where $A'_1$ denotes Duplicator's response).
Similar assumptions will be implicitly made throughout the proof.

\subtlt{Choice of $v_1$.}
Given colored graphs $G,H$ and a set $A\subset V(G)$, let $m(G,A;H)$
denote the number of $G\setminus A$-flaps $F$ such that $G[A\cup F]\cong H$.
Choice of $v_1$ is based on the following key observation:
Since $G\not\cong G'$, for some $H$ we have $m(G,A_1;H)\ne m(G',A'_1;H)$.
To be specific, suppose that the former number is larger.
Spoiler makes a composite move. In the first act he selects the union of
all $G\setminus A_1$-flaps $F$ contributing to $m(G,A_1;H)$.
The set selected by Duplicator in $G'$ obviously must contain a vertex $v'_1$
in a $G'\setminus A'$-flap $F'$ such that $G'[A'_1\cup F']\not\cong H'$.
In the second act Spoiler pebbles it. Whatever Duplicator's response $v_1$ is,
it belongs to a $G\setminus A_1$-flap $F$ such that $G[A_1\cup F]\cong H$.
Since $A_1\cup F=A_1\flap v_1$ and $A'_1\cup F'=A'_1\flap v'_1$,
Condition AB1 is ensured.
Condition AB2 follows from the fact that $A_1$ is a separator. 

However, it is not excluded that $G_1=G[A_1\cup F]$ and $G'_1=G'[A'_1\cup F']$
are disconnected. In this case we replace $G_1$ and $G'_1$ with their components
$\bar G_1$ and $\bar G'_1$ containing $F$ and $F'$ respectively. Since the rests
of $G_1$ and $G'_1$ are pebbled and hence are assumed to be isomorphic,
we have $\bar G_1\not\cong\bar G'_1$. Let $\bar A_1=A_1\cap V(\bar G_1)$
and $\bar A'_1=A'_1\cap V(\bar G'_1)$. Note that 
$\bar A_1$ and $\bar A'_1$ correspond to one another according to the pebbling
and that $\bar G_1=G[\bar A_1\flap v_1]$ 
and $\bar G'_1=G'[\bar A'_1\flap v'_1]$.
To not abuse the notation, we reset $G_1=\bar G_1$, $A_1=\bar A_1$
and $G'_1=\bar G'_1$, $A'_1=\bar A'_1$.

\subtlt{Choice of $B_1$.}
We make use of the induced tree decomposition of $G_1$, see Proposition \ref{prop:treedec}.1.
Spoiler pebbles a bag $B_1$ which is a separator of $G_1$, according to 
Proposition \ref{prop:treedec}.3. Duplicator must pebble $B'_1$ inside $G'_1$
to avoid Spoiler's fast win. This ensures Condition AB4. Since it is supposed that
$G_1$ has more than $2(k+1)$ vertices, $B_1\ne A_1$ and Condition AB5 follows.

\tlt{Phase $i+1$ following Phase $i$ of type AB.}
The game goes on $G_i$ and $G'_i$. By AB1 these graphs are non-isomorphic
and therefore $m(G_i,B_i;H)\ne m(G'_i,B'_i;H)$ for some colored graph $H$.
We first consider the case that there is a such $H$ with no colors from
$A_i\setminus B_i$. Then Phase $i+1$ has type AB.
Similarly to the choice of $v_1$, in the next composite move Spoiler
ensures pebbling $v_{i+1}\in F$ and $v'_{i+1}\in F'$, where $F$ and $F'$ are
flaps of $G_i\setminus B_i$ and $G'_i\setminus B'_i$ respectively,
so that exactly one of the graphs $G_i[B_i\cup F]$ and $G'_i[B'_i\cup F']$
is isomorphic to the $H$. To be specific, suppose that this is the former graph.
This means that $F\cap A_i=\emptyset$. We must also have $F'\cap A'_i=\emptyset$.
Otherwise $v'_{i+1}$ would be connected to a vertex in $A'_i$ within $G'_i\setminus B'_i$
while for $G$ the similar claim would be false, which would lead Spoiler to a fast win.
It easily follows that both $G_i[B_i\cup F]=G[B_i\flap v_{i+1}]$
and $G'_i[B'_i\cup F']=G'[B'_i\flap v'_{i+1}]$. 
Setting $A_{i+1}=B_i$ and $A'_{i+1}=B'_i$ ensures Conditions AB1 and AB2, the latter
by AB4 for Phase $i$. Condition AB3 is ensured similarly to Phase 1.
A set $B_{i+1}$ with AB4 and AB5 obeyed is pebbled also likewise.

Consider now the case that only choices of $H$ with colors from $A_i\setminus B_i$
are available. In this case one of the numbers $m(G_i,B_i;H)$ and $m(G'_i,B'_i;H)$
is equal to 1 and the other to 0. Then Phase $i+1$ has type ABC.

To be specific, suppose that $m(G_i,B_i;H)=1$ and let $F$ denote the flap
of $G_i\setminus B_i$ for which $G_i[B_i\cup F]\cong H$.
Set $A_{i+1}=A_i\cap F$ and $C_{i+1}=B_i$. 
Note that $B_i\cup F=A_{i+1}\betw C_{i+1}$.
A simple analysis shows that,
unless Spoiler has a fast win, we must have $A'_{i+1}\betw C'_{i+1}=B'_i\cup F'$
for some $G'_i\setminus B'_i$-flap $F'$.
We conclude that exactly one of $G_{i+1}$ and $G'_{i+1}$ is isomorphic to $H$,
which ensures Condition ABC1.
Condition ABC2 is true by AB4 for Phase $i$.

It is not excluded that $G_{i+1}$ is disconnected (if $G[B_i]$ was so).
Then, similarly to Phase 1, either Spoiler wins fast or we are able to
shrink $C_{i+1}$ so that the shrunken $G_{i+1}=G[A_{i+1}\betw C_{i+1}]$
becomes connected while ABC1 and ABC2 are preserved.

Spoiler pebbles a separator $B_{i+1}$ according to Propositions \ref{prop:treedec}.1
and \ref{prop:treedec}.3
to be a bag of the induced decomposition of $G_{i+1}$, say, $X_s\cap V(G_{i+1})$.
As Spoiler plays within $A_{i+1}\betw C_{i+1}$, Duplicator is forced to play
within $A'_{i+1}\betw C'_{i+1}$ and hence Condition ABC4 is obeyed.
By AB5 for Phase $i$, we have $A_{i+1}\subseteq X_r$ and $C_{i+1}\subseteq X_t$
for distinct $r$ and $t$. Since $G_{i+1}$ is connected 
and supposed to have more than $2(k+1)$
vertices, $B_{i+1}\notin\{A_{i+1},C_{i+1}\}$ and hence $s\notin\{r,t\}$.
Condition ABC5 is now not hard to infer from Proposition \ref{prop:treedec}.2.

\tlt{Phase $i+1$ following Phase $i$ of type ABC.}
As above, we are seeking for a colored graph $H$ such that $m(G_i,B_i;H)\ne m(G'_i,B'_i;H)$.
If a such $H$ exists with no color from $A_i\setminus B_i$ and $C_i\setminus B_i$,
Spoiler makes a composite move similarly to Phase 1 to ensure pebbling
$v_{i+1}$ in a $G_i\setminus B_i$-flap $F$ and $v'_{i+1}$ in 
a $G'_i\setminus B'_i$-flap $F'$ so that exactly one of the graphs $G_i[B_i\cup F]$ and 
$G'_i[B'_i\cup F']$ is isomorphic to the $H$. We can assume that both
$F\cap(A_i\cup C_i)$ and $F'\cap(A'_i\cup C'_i)$ are empty for else Spoiler wins fast.
It is easy to see that $B_i\cup F=B_i\flap v_{i+1}$
and $B'_i\cup F'=B'_i\flap v'_{i+1}$.
This allows Spoiler to perform Phase $i+1$ of type AB similarly to the above.

The next case we consider is that
there exists a choice of $H$ with colors from $A_i\setminus B_i$ but with
no color from $C_i\setminus B_i$ (or, what is treated symmetrically,
with colors from $C_i\setminus B_i$ but not from $A_i\setminus B_i$).
This case is also similar to the above. Unless Spoiler has a fast win,
there are a $G_i\setminus B_i$-flap $F$ and
a $G'_i\setminus B'_i$-flap $F'$ such that $F$ is the only flap
with $G_i[B_i\cup F]\cong H$ while $F'$ is determined by the condition that
$F'\cap A'_i$ and $F\cap A_i$ are under the same pebbles
(or all the same holds with $G$ and $G'$ interchanged). Moreover, we have
$G'_i[B'_i\cup F']\not\cong H$ and
$F'\cap C'_i=F\cap C_i=\emptyset$. It is not hard to see that
$B_i\cup F=A_{i+1}\betw C_{i+1}$ and $B'_i\cup F'=A'_{i+1}\betw C'_{i+1}$.
This allows Spoiler to perform Phase $i+1$ of type ABC in the same fashion as above.

There remains the case that only a choice of $H$ with colors both from $A_i\setminus B_i$ 
and from $C_i\setminus B_i$ is available. 
Again, unless Spoiler has a fast win,
there are a $G_i\setminus B_i$-flap $F$ and
a $G'_i\setminus B'_i$-flap $F'$ uniquely determined by the following conditions: 
For $F$ we have $G_i[B_i\cup F]\cong H$ and for $F'$ it is true that
$F'\cap (A'_i\cup C'_i)$ and $F\cap (A_i\cup C_i)$ are under the same pebbles
(or all the same holds with $G$ and $G'$ interchanged). Moreover, we have
$G'_i[B'_i\cup F']\not\cong H$. 
This case is most problematic because, following the same scenario
as above, for specification of $G_{i+1}$ we might be forced to keep up to
$3k+3$ pebbles on $A_i,B_i,C_i$, other $k+1$ pebbles might be needed to split
$G_{i+1}$ further, and it could not be excluded that we should keep up to
$4k+4$ pebbles to specify a $G_{i+2}$ and so on.

Spoiler has to make some extra efforts. It is easy to see that
$G_i[B_i\cup F]=G_i[B_i\flap u]$ for an arbitrary $u\in F\cap (A_i\cup C_i)$
and 
$G'_i[B'_i\cup F']=G'_i[B'_i\flap u']$ for the corresponding $u'\in F'\cap (A'_i\cup C'_i)$.
This allows Spoiler to force further play on these graphs, which we denote
by $\bar G_{i+1}$ and $\bar G'_{i+1}$.
We have $\bar G_{i+1}\not\cong\bar G'_{i+1}$ and 
$|V(\bar G_{i+1})|\le|V(G_i)|/2+k+1$ but we still need to shrink these graphs
to non-isomorphic $G_{i+1}$ and $G'_{i+1}$ either with 
$V(G_{i+1})\cap A_i=V(G'_{i+1})\cap A'_i=\emptyset$ or with
$V(G_{i+1})\cap C_i=V(G'_{i+1})\cap C'_i=\emptyset$.

We use Condition ABC5 for Phase $i$. By Proposition \ref{prop:treedec}.2,
$s$ is not on the path from $r$ to $t$. Let $q$ denote the vertex of $T$
such that $r,s,t$ are in pairwise distinct components of $T\setminus\{q\}$.
Spoiler pebbles the set of vertices $E=X_q\cap V(\bar G_{i+1})$
and Duplicator is forced to pebble a set $E'\subset V(\bar G'_{i+1})$.
By Proposition \ref{prop:treedec}.2, every flap of $\bar G_{i+1}\setminus E$
intersects at most one of the sets $A_i,B_i,C_i$.

Since $\bar G_{i+1}\not\cong\bar G'_{i+1}$, there is a colored graph $H$
with $m(\bar G_{i+1},E;H)\ne m(\bar G'_{i+1},E';H)$.
If a such $H$ exists only with colors of $E$, in a composite move Spoiler
ensures pebbling $v_{i+1}$ in a $\bar G_{i+1}\setminus E$-flap $F$
and $v'_{i+1}$ in a $\bar G'_{i+1}\setminus E'$-flap $F'$ such that
exactly one of the graphs $\bar G_{i+1}[E\cup F]$ and $\bar G'_{i+1}[E'\cup F']$
is isomorphic to $H$. Unless Spoiler wins fast, no vertex of $F$ and $F'$
is colored. Notice that $F$ is as well a $G_i\setminus E$-flap,
$\bar G_{i+1}[E\cup F]=G_i[E\cup F]$, and the same holds in $G'$. 
It is easy to see
that $E\cup F=E\flap v_{i+1}$ and $E'\cup F'=E'\flap v'_{i+1}$
in $G$ and $G'$ respectively. This allows Spoiler to proceed with Phase $i+1$
of type AB by setting $A_{i+1}=E$ and then pebbling $B_{i+1}$ as usually.

It only remains to consider the case that $H$ has colors exactly from one of
the sets $E\cup A_i$, $E\cup B_i$, and $E\cup C_i$. Without loss of generality,
suppose that $H$ has colors from $E\cup A_i$ and occurs (once) as an isomorphic
copy of $\bar G_{i+1}[E\cup F]$ for $F$ being a $\bar G_{i+1}\setminus E$-flap
(no such copy exists in $G'$). Unless Spoiler has a fast win, there is
a $\bar G'_{i+1}\setminus E'$-flap $F'$ with $F'\cap A'_i$ colored (i.e.\ pebbled)
identically to $F\cap A_i$ and $F'\cap B'_i=F'\cap C'_i=\emptyset$.
The graphs $\bar G_{i+1}[E\cup F]$ and $\bar G'_{i+1}[E'\cup F']$
are non-isomorphic because the former is isomorphic to $H$ while the latter is not.
Again, on the account of the observation that $F$ is also a flap of $G_i\setminus E$,
it is not hard to show
that $\bar G_{i+1}[E\cup F]=G[E\betw(F\cap A_i)]$ and, similarly,
$\bar G'_{i+1}[E'\cup F']=G'[E'\betw(F'\cap A'_i)]$.
Spoiler sets $A_{i+1}=A_i\cap F$, $C_{i+1}=E$ and concludes Phase $i+1$,
which is now of type ABC, as usually.

\tlt{Resources of the game.}
Spoiler needs the maximum number of pebbles in the case considered last.
This number can reach $4(k+1)$ if $A_i,B_i,C_i$ and $E$ all are bags of the decomposition
of $G$. Owing to Conditions AB2 and ABC2, the number of phases is smaller
than $\log(n-2k-2)+2\le\log n+3$ (for these estimates we assume that $n\ge4k+4$;
this restriction will be adsorbed in an additive term).
The number of rounds per phase is bounded by $2(k+1)$
(it is maximum also in the case consider last, where Spoiler
has to pebble first $E$ and only then $B_{i+1}$). 
Thus, the total number of rounds is less than $2(k+1)\log n+2k+9$.
Note that we do not have to add an extra $\log n$ term to count
the possibility that at some point Duplicator deviates from playing
on $G_i$ and $G'_i$ and Spoiler invokes a halving strategy. 
In this case we actually would have to add $\log|V(G_i)|$, which
is covered by the number of phases that could be played after Phase $i$
and, as such, is already included in the bound.

\section{Graph embeddings in orientable surfaces}\label{s:maps}

We here consider cellular embeddings of connected graphs in orientable surfaces of arbitrary genus 
using for them a standard combinatorial representation, see \cite[Section 3.2]{MTh}. 
A \emph{rotation system} $R=\langle G,T\rangle$ is a structure 
consisting of a graph $G$ and a ternary relation $T$ on $V(G)$
satisfying the following conditions:

\begin{enumerate}
\item
If $T(x,y,z)$, then $y$ and $z$ are in $\Gamma(x)$, the neighborhood of $x$ in $G$.
\item
For every $x$ the binary relation $T_x(y,z)=T(x,y,z)$ is a directed cycle
on $\Gamma(x)$ (i.e., for every $y$ there is exactly one $z$ such that $T_x(y,z)$
and for every $z$ there is exactly one $y$ such that $T_x(y,z)$).
\end{enumerate}

Geometrically, $T_x$ describes
the circular order in which the edges of $G$ incident to $x$ occur in the embedding
if we go around $x$ clockwise.

\begin{theorem}\label{thm:map}
Let $R=\langle G,T\rangle$ be a rotation system for a connected graph $G$
with $n$ vertices. We have $D^{5}(R)<3\log n+8$.
\end{theorem}

On the account of Corollary \ref{cor:TC}.2 this 
implies earlier results of Miller and Reif~\cite{MRe}.

\begin{corollary}\label{cor:map}
The isomorphism problem for rotation systems is in \ac1.
\end{corollary}

Miller and Reif give also a reduction of the planar graph isomorphism
to the isomorphism problem for rotation systems which is an \ac1 reduction provided
3-connected planar graphs are embeddable in plane in \ac1.
The latter is shown by Ramachandran and Reif~\cite{RRe}. 

\begin{corollary}\label{cor:planar}
The isomorphism problem for planar graphs is in \ac1.
\end{corollary}

In the rest of the section we prove Theorem \ref{thm:map}.
The proof is based on Equality \refeq{eq:dddc} and Proposition \ref{prop:game}.
Let $R=\langle G,T\rangle$ be a rotation system with $n$ vertices
and $R'=\langle G',T'\rangle$ be a non-isomorphic structure of the same signature.
We have to design a strategy for Spoiler
in the  Ehrenfeucht-Fra\"\i{}ss\'{e} game on $R$ and $R'$ allowing him to win with only
5 pebbles in less than $3\log n+8$ moves, whatever Duplicator's strategy.

The case that $R'$ is not a rotation system is simple. Spoiler needs just 4 moves
to show that $R'$, unlike $R$, does not fit the definition
(which has quantifier depth 4).
We will therefore suppose that $R'$ is a rotation system as well.

The main idea of the proof is to show that a rotation system admits a natural coordinatization
and that Duplicator must respect vertex coordinates. A coordinate system on $R=\langle G,T\rangle$
is determined by fixing its origins, namely, an ordered edge of $G$.
We first define \emph{local coordinates} on the neighborhood of a vertex $x$.
Fix $y\in\Gamma(x)$ and let $z$ be any vertex in $\Gamma(x)$. Then $c_{xy}(z)$
is defined to be the number of $z$ in the order of $T_x$ if we start counting from $c_{xy}(y)=0$.
In the global system of coordinates specified by an ordered pair of adjacent $a,b\in V(G)$,
each vertex $v\in V(G)$ receives coordinates $C_{ab}(v)$ defined as follows.
Given a path $P=a_0a_1a_2\ldots a_l$ from $a_0=a$ to $a_l=v$,
let $C_{ab}(v;P)=(c_1,\ldots,c_l)$ be a sequence of integers with 
$c_1=c_{ab}(a_1)$ and $c_i=c_{a_{i-1}a_{i-2}}(a_i)$ for $i\ge2$.
We define $C_{ab}(v)$ to be the lexicographically minimum $C_{ab}(v;P)$ over all $P$.
Note that $C_{ab}(v)$ has length $d(a,v)$.
By $P_v$ we will denote the path for which $C_{ab}(v)=C_{ab}(v;P_v)$.
One can say that $P_v$ is \emph{the extreme left shortest path from $a$ to $v$}.
Note that $P_v$ is reconstructible from $C_{ab}(v)$ and hence 
different vertices receive different coordinates.
The following observation enables a kind of the halving strategy.

\begin{lemma}\label{lem:split}
Let $a,b,v\in V(G)$ and $a',b',v'\in V(G')$, where $a$ and $b$ as well as
$a'$ and $b'$ are adjacent.
Assume that $d(a,v)=d(a',v')$ but $C_{ab}(v)\ne C_{a'b'}(v')$.
Furthermore, let $u$ and $u'$ lie on $P_v$ and $P_{v'}$ at the same distance
from $a$ and $a'$ respectively. Assume that $C_{ab}(u)=C_{a'b'}(u')$. 
Finally, let $w$ and $w'$ be predecessors of
$u$ and $u'$ on $P_v$ and $P_{v'}$ respectively.
Then $C_{uw}(v)\ne C_{u'w'}(v')$.
\end{lemma}

\begin{proof}
By definition, $C_{ab}(v)=C_{ab}(u)C_{uw}(v)$ and $C_{a'b'}(v')=C_{a'b'}(u')C_{u'w'}(v')$.
\end{proof}

\begin{lemma}\label{lem:coord}
Suppose that $a,b,v\in V(G)$ and $a',b',v'\in V(G')$
are pebbled coherently to the notation.
Assume that $a$ and $b$ as well as $a'$ and $b'$ are adjacent
and that $C_{ab}(v)\ne C_{a'b'}(v')$.
Then Spoiler is able to win with 5 pebbles in less than $3\log n+3$ moves.
\end{lemma}

\begin{proof}
Assume that $d(a,v)\ge2$.
If $d(a,v)\ne d(a',v')$, Spoiler wins in less than $\log n+1$ moves
by Lemma \ref{lem:halving}. If $d(a,v)=d(a',v')$, Spoiler applies a more
elaborated halving strategy.
Let $u$ be the vertex on $P_v$ with
$
d(a,u)=\lceil d(a,v)/2\rceil
$
and $u'$ be the corresponding vertex on $P_{v'}$.

\CASE 1{$C_{ab}(u)\ne C_{a'b'}(u')$.}
Without loss of generality assume that $C_{ab}(u)$ is lexicographically smaller
than $C_{a'b'}(u')$ (otherwise Spoiler moves in the other graph symmetrically).
Spoiler pebbles $u$. Denote Duplicator's response in $G'$ by $u^*$.
If $C_{ab}(u)=C_{a'b'}(u^*)$, then in our coordinate system $u^*$
is strictly on the left side to $P_{v'}$, the ``left most'' shortest path
from $a'$ to $v'$. It follows that $d(u^*,v')>d(u',v')=d(u,v)$ and
Spoiler wins fast by Lemma \ref{lem:halving}.
If $C_{ab}(u)\ne C_{a'b'}(u^*)$, then Spoiler has the same configuration as
at the beginning, with $u,u^*$ in place of $v,v'$, and with the distance
$d(a,u)$ twice reduced if compared to $d(a,v)$.
Then Spoiler does all the same once again.

\CASE 2{$C_{ab}(u)=C_{a'b'}(u')$.}
Spoiler pebbles $u$. If Duplicator responds with $u^*\ne u'$ then either
$d(a,u)\ne d(a',u)$ or $d(a,u)=d(a',u)$ but $C_{ab}(u)\ne C_{a'b'}(u^*)$
and Spoiler has a configuration similar to the beginning.
Assume therefore that $u^*=u'$.

Let $w$ and $w'$ be as in Lemma \ref{lem:split}.
Now Spoiler acts with $w,w'$ exactly in the same way as he just did with $u,u'$.
As a result, the players pebble vertices $\tilde w\in V(G)$ and
$\tilde w'\in V(G')$, where $\tilde w=w$ or $\tilde w'=w'$, with three possible outcomes:
\begin{enumerate}
\item
Some distances between the corresponding vertices in $G$ and $G'$ disagree.
\item
Spoiler achieves the same configuration as at the beginning with $\tilde w,\tilde w'$
in place of $v,v'$, where $d(a,\tilde w)<\lceil d(a,v)/2\rceil$.
\item
$\tilde w=w$ and $\tilde w'=w'$.
\end{enumerate}
In the first case Spoiler wins fast. In the third case Lemma \ref{lem:split}
applies and again Spoiler has the same configuration as at the beginning
with respect to new coordinate origins $(u,w)$ and $(u',w')$,
where $d(u,v)=\lfloor d(a,v)/2\rfloor$ is reduced.

In less than $\log d(a,v)+1$ iterations Spoiler forces a configuration as at the beginning
with $d(a,v)=1$ (we restore the initial notation), so it remains to consider this case.
Suppose that $d(a',v')=1$ as well. Now we have disagreement of local coordinates:
$c_{ab}(v)\ne c_{a'b'}(v')$. Keeping the pebbles on $a$ and $a'$, Spoiler restricts
play to the directed cycles $T_a$ and $T'_{a'}$ and wins with other 3 pebbles 
in less than $\log\deg a+1$
moves applying an analog of the strategy of Lemma \ref{lem:halving} for linear orders.

Each iteration takes at most 2 moves, which may be needed in Case 2.
Thus, Spoiler needs less than $2(\log\diam G+1)+(\log\Delta(G)+1)\le3\log n +3$
moves to win. The maximum number of pebbles is on the board in Case 2
(on $a,b,v,u,$ and $w$).
\end{proof}

Now we are ready to describe Spoiler's strategy in the game.
In the first two rounds he pebbles $a$ and $b$, arbitrary adjacent vertices in $G$.
Let Duplicator respond with adjacent $a'$ and $b'$ in $G'$.
If $G$ contains a vertex $v$ with coordinates $C_{ab}(v)$ different from
every $C_{a'b'}(v')$ in $G'$ or if $G'$ contains a vertex with
coordinates absent in $G$, then Spoiler pebbles it and wins by Lemma \ref{lem:coord}.
Suppose therefore that the coordinatization determines a matching between $V(G)$ and $V(G')$.
Given $x\in V(G)$, let $f(x)$ denote the vertex $x'\in V(G')$ with $C_{a'b'}(x')=C_{ab}(x)$.
If $f$ is not an isomorphism from $G$ to $G'$, then Spoiler pebbles two vertices
$u,v\in V(G)$ such that the pairs $u,v$ and $f(u),f(v)$ have different adjacency.
Not to lose immediately, Duplicator responds with a vertex having different coordinates
and again Lemma \ref{lem:coord} applies.
If $f$ is an isomorphism between $G$ and $G'$, then this map does not respect the
relations $T$ and $T'$ and Spoiler demonstrates this similarly.
The proof of Theorem \ref{thm:map} is complete.

\end{document}